\documentclass[sigconf]{acmart}

\usepackage{multirow}
\usepackage{textcomp}
\usepackage{todonotes}
\usepackage{framed}
\AtBeginDocument{%
  }

\setcopyright{acmcopyright}
\copyrightyear{2018}
\acmYear{2018}
\acmDOI{XXXXXXX.XXXXXXX}

\acmConference[WSESE 2024]{The 1st International Workshop on Methodological Issues with Empirical Studies in Software Engineering}{April 2024}{Lisbon, Portugal}

\acmPrice{15.00}
\acmISBN{978-1-4503-XXXX-X/18/06}

\begin{document}

\title{Are You a Real Software Engineer? Best Practices in Online Recruitment for Software Engineering Studies}

\author{Adam Alami}
\affiliation{%
  \institution{Aalborg University}
  \country{Denmark}
}

\author{Mansooreh Zahedi}
\affiliation{%
  \institution{University of Melbourne}
  \country{Australia}}

\author{Neil Ernst}
\affiliation{%
  \institution{University of Victoria}
  \country{Canada}}

\renewcommand{\shortauthors}{Alami et al.}

\begin{abstract}

  Online research platforms, such as Prolific, offer rapid access to diverse participant pools but also pose unique challenges in participant qualification and skill verification. Previous studies reported mixed outcomes and challenges in leveraging online platforms for the recruitment of qualified software engineers. Drawing from our experience in conducting three different studies using Prolific, we propose best practices for recruiting and screening participants to enhance the quality and relevance of both qualitative and quantitative software engineering (SE) research samples. We propose refined best practices for recruitment in SE research on Prolific. (1) Iterative and controlled prescreening, enabling focused and manageable assessment of submissions (2) task-oriented and targeted questions that assess technical skills, knowledge of basic SE concepts, and professional engagement. (3) AI detection to verify the authenticity of free-text responses. (4) Qualitative and manual assessment of responses, ensuring authenticity and relevance in participant answers (5) Additional layers of prescreening are necessary when necessary to collect data relevant to the topic of the study. (6) Fair or generous compensation post-qualification to incentivize genuine participation. By sharing our experiences and lessons learned, we contribute to the development of effective and rigorous methods for SE empirical research. particularly the ongoing effort to establish guidelines to ensure reliable data collection. These practices have the potential to transferability to other participant recruitment platforms.
  
\end{abstract}

\ccsdesc[500]{General and reference~Empirical studies}

\keywords{Empirical Software Engineering, Prolific, Participant Recruitment, Online Research Platforms}

\maketitle

\section{Introduction}\label{sec:introduction}

The recruitment of qualified software engineers remains a challenge \cite{ebert2022recruiting,rauf2022challenges,danilova2021you}. While the use of online research platforms such as Prolific\footnote{\url{https://www.prolific.co/}}, Mechanical Turk (MTurk)\footnote{\url{https://www.mturk.com/}}, and freelancing platforms like UpWork\footnote{\url{https://www.upwork.com/}} is increasing in the software engineering (SE) research community, reports are mixed. Some researchers report unsuccessful experience in recruiting qualified participants using Prolific \cite{ebert2022recruiting}, while others had relatively good experience using UpWork \cite{gutfleisch}. Overall, most researchers report challenging experiences using Prolific \cite{rauf2022challenges,reid2022software}.

Some of the challenges reported in some SE studies using Prolific are inherent to the platform itself. For example, the platform does not verify or assess self-reported skills. Motivated by the pay, participants may lie to access studies, as experienced by Ebert et al. \cite{ebert2022recruiting}. This is not peculiar to Prolific; MTurk also provides a restricted set of predefined screening parameters \cite{chandler2017lie}. Subsequently, researchers must prescreen their own subjects \cite{chandler2017lie}. Our experience shows that prescreening is critical to the success of the study and the quality of the data collected.

SE are increasingly resorting to online research platforms for data gathering \cite{reid2022software}. This is not unique to SE; social sciences researchers have been using online platforms for decades, and their usage continues to increase \cite{palan2018prolific}. For example, Bohannon reported that the number of published articles describing social science studies done with participants obtained via MTruk increased from 61 in 2011 to more than 1,200 in 2015 \cite{bohannon2016mechanical}. The effectiveness of these platforms is attractive to researchers \cite{palan2018prolific}. They provide access to a large pool of participants in a short period of time and can be inexpensive \cite{palan2018prolific}. Some even claim that access to a larger population is more representative of the population than traditional lab experiments with students \cite{bohannon2016mechanical,palan2018prolific}.

We opted for Prolific because it offers a wide-ranging demographic reach. It facilitates access to a diverse and global pool of participants from various backgrounds. Although other platforms exist (e.g., MTruk), Prolific provides reliability measures, making it a preferable choice over other platforms. The platform has stringent identity checks during the on-boarding process. In this paper, we reflect on and report our experience using Prolific for participant recruitment in three different studies \cite{alami2023antecedents,alami2023does,chase}.

Acknowledging the mixed outcomes and challenges reported in prior studies \cite{ebert2022recruiting,reid2022software,rauf2022challenges}, we recognize the need to establish common practices to overcome the challenges of participant qualification and skill verification. This is especially pertinent in the context of SE, where specific technical skills and experience are crucial to ensuring the quality of the data collected. Therefore, we propose to investigate:

\medskip

\noindent \textbf{RQ:} What are the best practices for recruiting and screening participants on platforms like Prolific to enhance the quality and relevance of qualitative and quantitative software engineering research samples?

\medskip

We share our experience, and we draw lessons on the best practices for recruiting and screening SE participants to enhance the quality and relevance of research samples in both qualitative and quantitative studies. Our report aims to contribute to the evolving landscape of online research methods, particularly in studies specific to SE. We contribute to the endeavor to establish best practices for reliable and representative data in SE empirical research.

\section{Related Work}\label{sec:related}

Reid et al. reported their experience recruiting software developers with Node.JS programming skills \cite{reid2022software}. They used a prescreening survey to validate their participants programming skills. Their experience highlights the significant limitation of solely relying on self-reported skills in Prolific to recruit programmers, which may lead to unqualified participants. To mitigate this, they recommend employing validation measures such as programming questions to evaluate participants skills \cite{reid2022software}.

Similarly, Ebert et al. reported their experience recruiting open source contributors in Prolific \cite{ebert2022recruiting}. Their findings underscore further the importance of prescreening in selecting participants with the requisite skills. Notably, participants had a tendency to provide misleading information to qualify for specific studies. To address this challenge, they suggest multiple rounds of validation to ensure the integrity of participants \cite{ebert2022recruiting}.

Danilova et al. proposed 16 questions using the Dillman pre-testing process to assess online participants programming skills \cite{danilova2021you}. The questions test programming language recognition, information sources, basic concepts, number formats, finding errors, algorithmic runtime, and program comprehension. They evaluated the questions using a survey instrument and found that programmers performed better than non-programmers in answering the questions. They tested the proposed questions with non-programmers, Clickworkers, and adversarial conditions and recommended six for use in prescreening for online studies. The questions they proposed are: ``Which of these websites do you most frequently use as an aid when programming?'' ``Choose the answer that best fits the description of a compiler’s function,'' ``Choose the answer that best fits the definition of a recursive function,'' ``Which of these values would be the most fitting for a Boolean?'' ``What is the parameter of the function?'' ``Please select the returned value of the pseudo code'' \cite{danilova2021you}. However, with the advent of large language model (LLM) tools like ChatGPT, which enable even non-programmers to swiftly respond to typical prescreening questions, we adopted a different strategy. We focused on task-oriented questions that required free-text responses. This approach enabled us to manually review each response, ensuring a thorough verification of the authenticity and qualifications of the participants.

\begin{table*}[t!]
    
    \footnotesize
    \caption{Summary of the three studies}
    \vspace{-0.3cm}
    \label{tbl:studies}
    
    \renewcommand\arraystretch{1.0}
  
    \begin{tabular}{lp{9cm}ccc}
    
    \toprule
    \textbf{Studies} & \textbf{Brief summary} & \multicolumn{3}{c}{\textbf{Recruitment Statistics}} \\ \cline{3-5}
    & & \textbf{Pre-screening} & \textbf{Eligible \& Invited} & \textbf{Final participants} \\ \hline
      
        \multirow{4}{*}{\emph{Study I}} & In this study, we investigated the antecedents of psychological safety within agile software development teams, building on the claim that agile practices are ineffective without it. It is a two-phase mixed-methods approach. The initial phase involved 18 exploratory interviews, followed by a survey study with 365 participants. & 914 & 436 & 365 \\ \hline
        
        \multirow{4}{*}{\emph{Study II}} & The study investigates the impact of psychological safety on agile teams' pursuit for software quality. The research aims to determine how this team's trait influences software quality outcomes. Utilizing a two-phase mixed-methods design, the study first conducted 20 exploratory interviews, followed by a survey with 423 participants. & 1000 & 480 & 423\\ \hline
        
        \multirow{2}{*}{\emph{Study III}} & In this interview study, we aimed to understand the factors that promote individual accountability among software engineers for outcomes such as software quality, security, and meeting project's deadlines.& 565 & 562 & 12\\
    
    \bottomrule
      
  \end{tabular}

\end{table*}

\section{Methods}\label{sec:methods}

In this section, we describe three studies that inform our best practices. We will refer to these studies as \emph{Study I} \cite{alami2023antecedents}, \emph{Study II} \cite{alami2023does}, and \emph{Study III} \cite{chase}. In the three studies, we used Prolific successfully to recruit for two separate surveys (N = 365 for \emph{Study I}, and N = 423 for \emph{Study II}), and twelve interviewees for \emph{Study III} (Tbl. \ref{tbl:interviewees}). Table \ref{tbl:studies} summarizes the studies and the numbers of participants in each phase of recruitment.

\begin{table*}[th!]
    
    \footnotesize
    \caption{Study I Prescreening Survey Questions}
    \vspace{-0.3cm}
    \label{tbl:prescreening}
    
    \renewcommand\arraystretch{1.0}
  
    \begin{tabular}{lp{10.5cm}p{5.5cm}}
    
    \toprule
      \textbf{\#} & \textbf{Prescreening Question} & \textbf{Measures}\\ \hline
      
        Q1 & Are you currently working in an agile software development team? &  Multiple choice (Yes and No)\\ \hline
        Q2 & What is your role in your software development team? & Multiple choice (e.g., Software engineer, QA, Tech Lead, Product Owner, Other, etc.)\\ \hline
        Q3 & What agile method do you use in your team? & Multiple choice (e.g., Scrum, XP, Kanban, etc.)\\ \hline
        Q4 & Is this statement true or false: Agile is a family of software development methods, inspired by the ``Agile Manifesto.'' & Multiple choice (True and False) \\ \hline
        \multirow{4}{*}{Q5} & We use ISO/IEC 25010 definition of software quality in this study, which states: ``[Software quality is] the degree to which the system satisfies the stated and implied needs of its various stakeholders, and thus provides value.'' This ISO model also covers some non-functional characteristics, mainly, ``performance'', ``compatibility'', ``usability'', ``reliability'', ``security'', ``maintainability'', and ``portability.'' Do you agree with this definition? & \multirow{4}{5cm}{5-level Likert scale (i.e., strongly disagree to strongly agree)} \\ \hline
        Q6 & Please, discuss your response below: & Free text \\
    
    \bottomrule
      
  \end{tabular}
  \vspace{-0.3cm}

\end{table*}

\subsection*{Study I}

In this study, we sought to recruit quality assurance (QA) professionals and software engineers working in agile software development teams. Aware of the potential limitations of self-reported screening data from Prolific, we commenced our recruitment with a thorough prescreening exercise. The prescreening process was iterative. We launched daily prescreening surveys with a limit of up to 50 respondents. We capped the number of respondents to allow us to scrutinize the prescreening data closely and manually. The limit of 50 respondents was chosen based on the researcher capacity to conduct evaluations on a daily basis. Table \ref{tbl:prescreening} documents our prescreening questions. All the questions were mandatory. The prescreening process took place between May 2nd and May 26th, 2022. For each iteration of the prescreening, we conducted a five-step elimination process:

\noindent \textbf{Step 1:} We eliminated all entries with a ``No'' response to Q1.

\noindent \textbf{Step 2:} We eliminated all entries with an ``Other'' response to Q2, and the free text response was not a software development or QA-related role.

\noindent \textbf{Step 3:} We eliminated all entries with an ``Other'' response to Q3, and the free text response was not an agile method.

\noindent \textbf{Step 4:} We eliminated all entries with ``False'' responses to Q4. Responding ``False'' to this question is an indication that the participant is either not genuine and/or unaware of agile software development.

\noindent \textbf{Step 5:} We examined the comments in the free text of Q6 to assess their quality manually. We eliminated entries that we deemed to be ambiguous, inauthentic, uncomprehensible, or indicating low efforts made by the participant. The quality of the comments varied extremely, from text that copied and pasted the text of question Q5 to blunt text admitting a lack of knowledge on the topic. However, genuine comments have shown depth and familiarity with the topic; for example, a respondent commented, ``Software quality cannot be solely described by satisfying the various stakeholders needs.''

The prescreening phase attracted 914 respondents. Then, the qualifying process (steps discussed above) yielded a reliable sample of 436 potential participants; the percentage of genuine participants in the prescreening sample was approximately 47.70\%.

In this prescreening process, we used several techniques that have allowed us to achieve a highly qualified sample: \emph{iterative and controlled prescreening}, \emph{comprehensive and targeted screening questions}, and \emph{qualitative assessment}.

\noindent \textbf{Iterative and controlled prescreening.} The decision to conduct daily prescreening surveys with a limited number of respondents (up to 50) allowed us to manage the data volume and thoroughly evaluate the small volume of entries. This approach allowed for detailed manual scrutiny of responses, ensuring that each entry received adequate attention. The low volume also allowed us a more nuanced scrutiny of our participants, like their understanding of agile methodologies (i.e., Q4) and perceptions of software quality (i.e., Q5 \& Q6). This layered approach helps in progressively narrowing down the most genuine participants.

\noindent \textbf{Comprehensive and targeted screening questions.} Although we did not ask for programming skills-related questions because QA sometimes does not have programming skills, we used a question that not only tested the skills of QA and software engineers but also their professional engagement. Asking QA professionals and software engineers to assess the ISO/IEC 25010 definition of software quality in a prescreening process is not only a test of their specific knowledge but also a measure of their critical thinking, practical application skills, professional engagement, and overall experience in the industry. Such a task can effectively differentiate genuine, well-qualified professionals from those with superficial understanding or who are inauthentic or unqualified. For example, unqualified participants were unable to supplement responses to support their disagreement or agreement with the ISO/IEC 25010 definition of software quality. Their responses were either uncomprehensive or, in some instances, a random copy and paste from a Google search.

\noindent \textbf{Qualitative assessment.} The qualitative analysis is in step 5, where responses were evaluated based on the quality and rationality of the comments provided. These steps allowed us to assess the participants' depth of knowledge and engagement with the subject matter, beyond mere agreement or disagreement.

\subsection*{Study II}

We used a similar approach to the previous study (i.e., \emph{Study I}) successfully. Similarly, over a period of 20 days, we collected up to 50 entries in a prescreening survey using same questions. This time, the prescreening phase attracted 1,000 respondents. Then, the qualifying process (steps discussed above) yielded a reliable sample of 480 qualified participants; the percentage of genuine participants in the prescreening sample was 48\%. Using similar process across both studies demonstrated the reliability of our prescreening approach. This suggests that the method is robust and can be effectively replicated in similar studies. We also become more confident on the crucial role of the qualitative assessment step. It reinforced the value of incorporating a qualitative review when participant expertise and engagement are critical.

\subsection*{Study III}

\begin{table*}[t!]
    
    \footnotesize
    \caption{Study III Prescreening Survey Questions}
    \vspace{-0.3cm}
    \label{tbl:prescreening2}
    
    \renewcommand\arraystretch{1.6}
  
    \begin{tabular}{lp{10.5cm}p{5.5cm}}
    
    \toprule
      \textbf{\#} & \textbf{Prescreening Question} & \textbf{Instrument}\\ \hline
      
        Q1 & What is your role in your software development team? & Multiple choice (e.g., Software engineer, QA, Tech Lead, Other, etc.)\\ \hline
        Q2 & What software development process or method do you use in your team? & Scrum, XP, Kanban, Pla-driven (e.g., Waterfall, Prince2, etc.)\\ \hline
        Q3 & How many years of experience do you have working as a software developer or quality assurance analyst or software engineer? & Less tan 5 years, or More than 5 years\\\hline
        Q4 & How do you identify your gender? & Male, Female, etc. \\\hline
        Q5 & Is this statement true or false: Agile is a family of software development methods, inspired by the ``Agile Manifesto.'' & Multiple choice (True and False) \\ \hline
        \multirow{6}{*}{Q6} & Please write a function in your preferred programming language that takes a string as input and performs one of the following tasks (AI detector will be used to verify your answer): & Free text \\ 
        &  1. Reverse the String: Reverse the input string. & \\
        &  2. Count Vowels: Count the number of vowels (A, E, I, O, U) in the string. & \\
        &  3. Remove Duplicates: Remove any duplicate characters from the string, keeping only the first occurrence of each character. & \\
        &  4. Uppercase Conversion: Convert all characters in the string to uppercase. & \\\hline
        Q7 & Describe and discuss the differences between unit testing, system testing, and end-to-end testing? & Free text \\\hline
        Q8 & Please describe a specific instance where you, contributed to the success of a software development project. What challenges did you face, and how did your involvement impact the project's outcome? & Free text \\
    
    \bottomrule
      
  \end{tabular}
  \vspace{-0.3cm}

\end{table*}

In this study, we sought to recruit only software engineers for interviews. Capitalizing on lessons we learned from \emph{Study I \& II}, we deployed similar techniques: \emph{iterative and controlled prescreening}, \emph{comprehensive and targeted screening questions}, \emph{qualitative assessment}. In this instance, we included programming skills while still using a task-oriented approach. Aware of the potential of using LMM tools by participants, we opted for task-oriented questions (i.e., Q6, Q7, and Q8) to qualitatively assess the authenticity of the responses. In addition, we used ChatGPT for AI detection for questions Q6, Q7, and Q8.

After filtering non-software engineer roles, i.e., Q1, we scrutinized Q6, Q7, and Q8 answers. First, all answers to Q6 were screened for AI detection. After evaluating the answer manually, we used ChatGPT to test for AI-generated text and responses. Two answers did not pass the detection, and one answer was an incomprehensible text. Then, we evaluated the responses to Q7 and Q8 the same way, i.e., in a two-step process of manual verification followed by AI detection. We received 200 entries, and 197 qualified. The unqualified submission rate was significantly low in this prescreening; however, the dropout rate was high 30\%. This is highly likely because unqualified participants opted not to complete or submit their entry at Q6, fearing rejection of payment if they did not pass the AI test. A rejection of payment looks negatively in participants' profiles as their ``completion rate'' will go down and the risk of not being selected for future studies becomes a possibility. Another explanation is that the study is interviews instead of a survey, which may have deterred cheaters from participation. Lastly, in the first page of the prescreening survey, we introduced the study and disclosed that AI-detection shall be used when relevant.

We wanted interviewees to share their experiences of feeling accountable towards software engineering outcomes such as software quality, code quality, and software security. We aimed to recruit participants with different levels of accountability in their respective teams; hence, we carried out a second prescreening survey to collect data relevant to the concept of accountability. The purpose of this additional pre-screening is to collect data on how accountability mechanisms shape the participant's work environment, which we did not have in previous pre-screening data. Table \ref{tbl:prescreening3} documents some of the second prescreening survey questions.

We combined \emph{Study I} and the new pre-screening \emph{Study III}, to curate a total of 562 qualified participants. We opted to leverage \emph{Study I} participants, because they were vetted and both studies samples use software engineers. We invited the 562 to the second pre-screening. We received 159 responses, and then we chose twenty participants to participate in the interviews. We used a comprehensive set of criteria to select our interviewees (Tbl. \ref{tbl:interviewees}), spanning factors such as country of residence, age, experience, role, gender, education, software development method (e.g., agile, plan-driven, etc.), team size, project scope (i.e., what the SE team is developing), type of development (e.g., custom or maintenance), in-house development or outsourcing environments, and accountability practice within the participant team and organization. The number of participants  was determined by the point of thematic saturation, which was achieved at twelve. All participants attended the interview as scheduled and showed a high level of professionalism and experience. All our interview transcripts and interview guide are available \href{https://doi.org/10.5281/zenodo.10105278}{here.}\footnote{\url{https://doi.org/10.5281/zenodo.10105278}}

Building up on \emph{Study I} and \emph{Study II} learning, we propose additional techniques for prescreening: \emph{AI detection}, \emph{task-oriented questions}, and \emph{additional layer of prescreening based on the study's scope}.

\noindent \textbf{AI detection}. By incorporating AI detection in \emph{Study III}, we addressed the challenge of potential AI-generated responses, ensuring the authenticity of participants' submissions. In this study, we combined skills evaluation questions (i.e., Q6 and Q7), but we also sought insights into the respondents' practical experience. Our Q8 was strategically designed to elicit detailed, experiential responses, which has helped to distinguish genuine software engineers from those who are not. The question’s focus on real-world experience, problem-solving, and personal contribution makes it a powerful tool for identifying authentic and competent participants for our study.

\noindent \textbf{Task-oriented questions}. After three studies using task-oriented questions, we learned they effectively evaluated the practical skills and experiential knowledge of the participants. Combining questions to test technical expertise, knowledge in real-world situations, and critical thinking has yielded qualified and authentic samples.

\noindent \textbf{Additional layer of prescreening}. \emph{Study III} was about how software engineers feel accountable for some of their teams' outcomes. We needed further data to select a diverse and representative sample. Questions Q1-Q8 (Tbl. \ref{tbl:prescreening3}) aligned our selection closely with the study's objectives and also showed us potential participants rich insights into the topic of accountability.

\begin{table*}[t!]
    
    \footnotesize
    \caption{Study III Second Prescreening Survey Questions}
    \vspace{-0.3cm}
    \label{tbl:prescreening3}
    
    \renewcommand\arraystretch{1.6}
  
    \begin{tabular}{lp{10.5cm}p{5.5cm}}
    
    \toprule
      \textbf{\#} & \textbf{Prescreening Question} & \textbf{Instrument}\\ \hline
      
        Q1 & Please, briefly describe the scope of the project you are currently working on & Free text\\ \hline
        Q2 & Please provide a brief description of the most critical outcomes for which your team is held accountable. These outcomes should reflect specific goals or results that your team aims to achieve. For example, timely delivery of software releases, ensuring software security compliance, meeting Sprints' goals, etc. & Free text\\ \hline
        Q3 & Please provide a brief description of the most critical or important outcomes for which you are individually held accountable in your team. These outcomes should reflect specific goals or results that your team expects from you. For example, code quality, ensuring software security compliance, meeting deadlines, etc. & Free text \\\hline
        Q4 & In our team, we employ formal rules, guidelines, (e.g., coding standards) and established expectations (e.g., KPIs) to assess individual performance concerning the outcomes mentioned in the previous question. & 5-level Likert scale \\\hline
        Q5 & In our team, we employ tools and processes (e.g., code review, or annual performance review) to assess individual performance concerning the outcomes mentioned in the previous question. & 5-level Likert scale \\ \hline
        Q6 &  In our team, we employ informal rules, guidelines (e.g., transparent communication, and honesty), and expectations (e.g., helping each others) to assess individual performance concerning the outcomes mentioned in the previous question. & 5-level Likert scale \\\hline
        Q7 &  In our team, individuals feel accountable for their actions, decisions, and contributions related to the outcomes I mentioned in the previous question. & 5-level Likert scale \\\hline
        Q8 &  I feel accountable for my actions, decisions, and contributions related to the outcomes I mentioned in the previous question. & 5-level Likert scale \\
    
    \bottomrule
      
  \end{tabular}
   \vspace{-0.3cm}

\end{table*}

\begin{table*}[t!]

  \begin{center}
    \footnotesize
    \caption{Study III interviewees' characteristics.}
    \vspace{-0.3cm}
    \label{tbl:interviewees}
    \renewcommand\arraystretch{0.80}
        
    \begin{tabular}{l|p{2.5cm}|c|c|p{3cm}|p{3.5cm}|c|p{1.1cm}}
      \hline
      \textbf{\#} & \textbf{Role} & \textbf{Exp.} & \textbf{Method} & \textbf{Project type} & \textbf{Industry sector} & \textbf{Gender} & \textbf{Country}\\
      \hline
    
        P1 & Software Engineer & 3-5 years & DevOps & Media data platform & Information Technology services & Male & Germany\\
        P2 & Software Engineer & 3-5 years & Hybrid & Utility software & Information Technology services & Male & UK\\
        P3 & Software developer & 9-11 years & Scrum & Robotics software & Robotics manufacturing & Male & USA\\
        P4 & Software developer & 9-11 years & Hybrid & Business intelligence & Information Technology services & Male & Italy\\
        P5 & Sr. software engineer & 6-8 years & Scrum & Online market place & Information Technology services & Non-binary / third gender & Germany\\
        P6 & Sr. software engineer & $>$12 years & Kanban & Infrastructure migration & Banking services & Male & Canada\\
        P7 & Software developer & $<$3 years & Hybrid & Embedded software & Information Technology services & Male & France\\
        P8 & Sr. software engineer & $>$12 years & Scrum & Data migration & Information Technology services & Female & India\\
        P9 & Sr. software engineer & 6-8 years & Scrum & CRM software & Information Technology services & Male & Serbia\\
        P10 & Sr. software engineer & 9-11 years & Hybrid & Data management & Global software vendor & Male & Canada\\
        P11 & Software developer & 3-5 years & Scrum & Telecommunication software & Global software vendor & Male & UK\\
        P12 & Software engineer & $<$3 years & DevOps & Process automation & Information Technology services & Female & India\\

     \bottomrule
     
    \end{tabular}
   
  \end{center}
   \vspace{-0.3cm}
   
\end{table*}

\noindent \textbf{Compensation}. For \emph{Study I} and \emph{Study II}, we paid \textsterling0.50 for each submission in the prescreening. For \emph{Study III}, we paid \textsterling0.75 for the first prescreening and \textsterling1.25 for the second prescreening and \textsterling30 to each interviewee. We learned that irrespective of the pay of the prescreening, unqualified participants still submit entries. \emph{Study I} and \emph{Study II} received submissions for unqualified participants, 47.70\% and 48\% respectively. However, once qualified, the pay should be fair or even generous to attract software engineers to continues in subsequent phases of the studies, beyond prescreening.

Our experiences show that Prolific can be utilized effectively for recruiting qualified and genuine software engineering professionals, ensuring the collection of high-quality data for their studies. However, the prescreening process must be rigorous, and meticulously designed and executed.

\section{Best Practices}\label{sec:findings}

Prolific employs various techniques to manage and regulate accounts, yet the information submitted by platform prospects remains unreliable. To ensure security and prevent fraudulent activities, the platform has several measures in place. These include verifying each account using a non-VOIP phone number, allowing only one unique number per account. They also restrict signups based on IP and ISP, blocking low-trustworthy ones. Additionally, they limit the number of accounts that can use the same IP address and machine to prevent duplicate accounts. To further prevent misuse, they also limit the number of unique IPs per ``HIT'' (study) and require a unique PayPal account per participant.

The platform also used identification verification as part of onboarding participants. Prior to establishing their accounts, participants identities are vetted. They have to submit a copy of their passport, which is used by Prolific for verification. This vetting process does not deter participants from reporting misleading information to access studies, incentivized by the monetary rewards.

Drawing from the experiences described in Sect. \ref{sec:methods}, we propose the following best practices for software engineering research recruitment on Prolific: (1) \emph{iterative and controlled prescreening}, (2) \emph{task-oriented and targeted questions}, (3) \emph{AI detection}, (4) \emph{qualitative assessment of responses}, (5) \emph{additional layer of prescreening}, and (6) \emph{fair/generous pay after qualification.}

\paragraph*{\textbf{(1) Iterative and controlled prescreening.}} Conducting periodic prescreening surveys with a limited number of respondents allows for a focused and manageable assessment process. This can be run over short cycles to allow for a low volume of entries. This approach ensures detailed manual scrutiny of each response and maintains the quality of the assessment process.

\paragraph*{\textbf{(2) Task-oriented and targeted questions.}} Develop questions that go beyond mere theoretical understanding, as participants must apply their skills to solve specific problems or discuss complex scenarios. We recommend developing prescreening questions that not only assess technical skills but also gauge professional engagement and critical thinking. Questions should be relevant to the specific roles and specific skills they should have, e.g., programming skills. We found that task-oriented questions using free text are efficient in scrutinizing authenticity and skills. We also found that combining skills questions with questions to test professional engagement and critical thinking (e.g., evaluating the ISO/IEC 25010 definition for quality and contributing to a project's challenge) is rewarding. This technique showed that the answers revealed the depth of the participant's understanding. Genuine participants provided detailed, coherent, and contextually relevant answers, while those with superficial knowledge have struggled to articulate well-thought-out responses. Q8 (in Tbl. \ref{tbl:prescreening2}) and the second prescreening (Q1-Q8 in Tbl. \ref{tbl:prescreening3}) of \emph{Study III} have also shown that responses offer clearer and more in-depth insights into the participant's real-world expertise and competencies.

\paragraph*{\textbf{(3) AI detection.}} All free-text responses should be AI-detected. For example, participants can easily prompt an LLM tool to generate responses. In \emph{Study III} prescreening, three participants submitted AI-generated content (for Q6 in Tbl. \ref{tbl:prescreening2}). We used ChatGPT to detect AI-generated content using this prompt: ``assess whether this content is AI-generated:'' followed by the response. While AI-detection tools are increasing, their efficiency is still questionable. However, Yan et al. claim that after fine-tuning GPT-3, they achieve a ``99\%'' accuracy rate for AI-generated text detection \cite{yan2023detection}. Akram, on the other hand, studied several tools (e.g., ``GPTZero,'' ``Originality'' etc.) \cite{akram2023empirical}. They reported accuracy rates between 55.29\% and 97.0\% \cite{akram2023empirical}. As AI technologies evolve, and becoming more sophisticated in mimicking human outputs, this practice will remain a challenge.

\paragraph*{\textbf{(4) Qualitative assessment of responses.}} Once relevant responses have been verified for AI generation, we recommend manually reading and verifying the submitted text. This assessment goes beyond just verifying the authenticity of responses; it evaluates the quality, coherence, and relevance of the content. This approach helped us identify participants who possess not only the necessary technical skills but also the ability to articulate comprehensively their ideas and experiences. This is critical to ensuring that the research sample comprises well-qualified and insightful individuals. Such a thorough evaluation is key to gathering high-quality data that can significantly contribute to the depth and validity of the research findings. This is possible and even efficient when the prescreening is carried out in small cycles with low volumes. Practices (3) and (4) are also relevant to surveys' open-questions and online forms.

\paragraph*{\textbf{(5) Additional layer of prescreening.}} Implement an additional layer of prescreening when necessary to collect data on the specific aspects of the study, like accountability in software engineering in the case of \emph{Study III}. This helped in collecting additional data, which has provided further insights about the participants relevant to the study’s objectives. For interviews, we opted to collect insights from the second prescreening survey to understand the professional context of the participants, their roles and responsibilities, and the dynamics of accountability within their teams. We asked questions about the project's scope, team-level, and individual-level feeling of accountability (Tbl. \ref{tbl:prescreening3}). This information was invaluable in selecting participants who could provide rich, contextual, and relevant contributions to the study.

\paragraph*{\textbf{(6) Fair/generous pay after qualification.}} We do not recommend paying more than the Prolific recommendation for the first prescreening, i.e., \textsterling9.00 per hour. This is because the survey will attract unqualified participants. We learned that software engineers and other SE professionals know and are willing to accept low pay in the prescreening, anticipating a fair or generous pay in the main study. We used Danish standards as a point of reference to pay for the interviews. In Denmark, software engineers make on average DKK 318 an hour, equivalent to \textsterling37. However, we felt that we could have paid more to attract more participants. We invited 20, but only twelve accepted. For example, Gutfleisch et al. paid \$105 per interview and attracted 25 participants \cite{gutfleisch}.

\section{Conclusion}\label{sec:conclusion}

Recruiting genuine participants for SE research presents unique challenges, particularly in verifying the authenticity of participant's skills. Based on our previous studies, we draw a set of best practices for SE participants recruitment. Admittedly, our approach is labor intensive; however, it has yielded good results. While our suggested best practices may yield robust samples, the online research channels for recruitment and the technological landscape are ever evolving. Our community, should regularly review and refine prescreening and recruitment techniques based on shared experiences. Continuously learning from experiences should help enhancing the quality of recruitment process for future studies.

\bibliographystyle{ACM-Reference-Format}
\bibliography{references}

\end{document}